\documentclass[preprint]{JHEP3}

\usepackage{epsfig}
\usepackage{amsmath}
\usepackage{graphicx,psfrag}
\usepackage{dcolumn}
\usepackage{bm}
\usepackage{slashed}

\newcommand{\beq}{\begin{equation}}
\newcommand{\eeq}{\end{equation}}
\newcommand{\bea}{\begin{eqnarray}}
\newcommand{\eea}{\end{eqnarray}}
\newcommand{\hf} {\frac{1}{2}}

\newcommand{\nn}{\nonumber\\}

\newcommand\fig[1]     {Fig.\,{\ref{#1}}}

\def\Tr{{\rm Tr}}

\def\eq#1{(\ref{#1})}
\def\s0#1#2{\mbox{\small{$ \frac{#1}{#2} $}}}
\def\0#1#2{\frac{#1}{#2}}

\def\mr#1{{\mathrm{#1}}}

\title{Functional renormalization group 
with a compactly supported smooth regulator function}

\author{I. N\'andori,\\
MTA-DE Particle Physics Research Group, P.O.Box 51, H-4001 Debrecen, Hungary, \\
Institute of Nuclear Research, P.O.Box 51, H-4001 Debrecen, Hungary} 

\abstract{
The functional renormalization group equation with a compactly supported smooth 
(CSS) regulator function is considered. It is demonstrated that in an appropriate 
limit the CSS regulator recovers the optimized one and it has derivatives of all 
orders. The more generalized form of the CSS regulator is shown to reduce to
all major type of regulator functions (exponential, power-law) in appropriate limits.
The CSS regulator function is tested by studying the critical behavior of the 
bosonized two-dimensional quantum electrodynamics in the local potential 
approximation and the sine-Gordon scalar theory for $d<2$ dimensions beyond 
the local potential approximation. It is shown that a similar 
smoothing problem in nuclear physics has already been solved by introducing the 
so called Salamon-Vertse potential which can be related to the CSS regulator.
}

\keywords{functional renormalization group, optimization}

\preprint{}

\begin{document}

\section{Introduction}
\label{sec_intro}
The functional renormalization group (RG) method has been developed in 
order perform renormalization non-perturbatively, i.e. to determine the underlying 
exact low-energy effective theory without using perturbative treatments 
\cite{WeHo1973,Po1984,We1993,Mo1994,internal}. The functional RG equation 
in its most general form (for scalar fields) \cite{We1993}
\bea
\label{erg}
k \partial_k \Gamma_k [\phi] = \hf  \Tr \left[
(k\partial_k R_k) / (\Gamma_k^{(2)}[\phi] + R_k)
\right]
\eea
is derived for the blocked effective action $\Gamma_k$ which interpolates between 
the bare $\Gamma_{k\to \Lambda} = S$ and the full quantum effective action 
$\Gamma_{k\to 0}=\Gamma$ where $k$ is the running momentum scale. The second 
functional derivative of the blocked action is represented by $\Gamma_k^{(2)}$ and
the trace Tr stands for the momentum integration. $R_k$ is an appropriately 
chosen regulator function which fulfills the following requirements, $R_k(p\to 0)>0$, 
$R_{k\to 0}( p)=0$ and $R_{k\to \Lambda}( p)=\infty$. Since the RG equations are 
functional partial differential equations it is not possible to solve them in general, 
hence, approximations are required. One of the commonly used systematic 
approximation is the truncated derivative (i.e. gradient) expansion where the 
blocked action is expanded in powers of the derivative of the field,
\beq
\label{deriv}
\Gamma_k [\phi] = \int_x \left[V_k(\phi) 
+ Z_k(\phi) \hf (\partial_{\mu} \phi)^2 + ... \right].  
\eeq 
In the local potential approximation (LPA), i.e. in the leading order of 
the derivative expansion \eq{deriv}, higher derivative terms are neglected 
and the wave-function renormalization is set equal to constant, 
i.e. $Z_k \equiv 1$. The solution of the RG equations sometimes requires 
further approximations, e.g. the potential can be expanded in powers of the 
field variable. Since the approximated RG flow depends on the choice of the 
regulator function, i.e. on the renormalization scheme,  the physical results 
(such as fixed points, critical exponents) could become scheme-dependent.

Therefore, a general issue is the comparison of results obtained by various 
RG schemes (i.e. various types of regulator functions) \cite{opt_rg,litim_o(n),
opt_func,Ro2010,Mo2005,qed2,scheme,scheme_sg,sg_KTB,minimal_sens,reuter}. 
In order to optimize the scheme-dependence and to increase the convergence 
of the truncated flow (expanded in powers of the field variable), a general 
optimization procedure has already been worked out \cite{opt_rg,opt_func} 
and the link between the optimal convergence and global stability of the flows 
was also discussed. Optimization scenarios has also been discussed in detail 
in \cite{Ro2010}. Moreover, optimization through the principle of minimal 
sensitivity were also considered \cite{minimal_sens}. In the leading order of 
the derivative expansion \eq{deriv}, i.e. in LPA, an explicit form for the optimized 
(in a sense of \cite{opt_func}) regulator was provided \cite{opt_rg} but it was also 
shown that this simple form of the optimized regulator does not support a derivative 
expansion beyond second order \cite{Ro2010, Mo2005,opt_rg,opt_func}. The 
optimized regulator is a function of class $C^{0}$ with compact support thus it 
is a continuous function and it has a finite range but it is not differentiable. It was 
argued \cite{opt_rg,opt_func} that beyond LPA a solution to the general criterion 
for optimization (see Eq.(5.10) of \cite{opt_func}) has to meet the necessary 
condition of differentiability to the given order.

In this work we give an example for a regulator function of class $C^{\infty}$ 
(it has derivatives of all orders, i.e. it is a smooth function) with compact support. 
We show that in an appropriate limit it recovers the optimized regulator (optimized 
in a sense of \cite{opt_rg,opt_func}). Moreover, its generalized form can be 
considered as a prototype regulator which reduces to all major type of regulator 
functions (exponential, power-law) in appropriate limits. Finally, this regulator 
function is tested by studying the critical behavior of the bosonized two-dimensional 
quantum electrodynamics (QED$_2$) in LPA and the sine-Gordon scalar theory for 
$d<2$ dimensions beyond LPA.

\section{Regulator functions}
\label{sec_regulator}

A large variety of regulator functions has already been discussed in the 
literature by introducing its dimensionless form
\bea
R_k( p) = p^2 r(y),
\hskip 0.5cm
y=p^2/k^2
\eea
where $r(y)$ is dimensionless. For example, one of the simplest regulator 
function is the sharp-cutoff regulator
\beq
\label{sharp}
r_{\mr{sharp}}(y) = \frac{1}{\theta(y-1)} -1
\eeq 
where $\theta(y)$ is the Heaviside step function. The sharp-cutoff regulator 
has the advantage that the 
momentum integral in \eq{erg} can be performed analytically in the LPA. The 
corresponding RG equation is the Wegner-Houghton RG \cite{WeHo1973}. 
Its disadvantage is that it confronts to the derivative expansion, i.e. higher 
order terms (beyond LPA) cannot be evaluated unambiguously. 

The compatibility with the derivetive expansion can be fulfilled by e.g. using an 
exponential type regulator function such as \cite{We1993}
\beq
\label{exp}
r_{\mr{exp}}(y) = \frac{c}{\exp\left(c_2 y^b \right) -1}
\eeq 
with $b\geq 1$ and $c=1$ is a typical choice. The parameter $c_2$ can be 
chosen as e.g. $c_2 =\ln(2)$. Let us note, the exponential regulator with 
$c\neq 0$, $c_2 \neq \ln(2)$ has also been discussed in \cite{minimal_sens} 
using optimization through the principle of minimal sensitivity. Other 
exponential type regulators like, $r_{\mr{mexp}} = b/(\exp(c y) -1)$ with 
$c=\ln(1+b)$, $r_{\mr{mod}} = 1/(\exp(c [y + (b-1)y^b]/b) -1)$ with $c = \ln(2)$, 
$r_{\mr{mix}} = 1/(\exp(b [y^a - y^{-a}]/2a) -1)$ with $a \geq 0$ or 
$r_{\mr{step}} = (2b-2) y^{b-2}/(b[\exp(c y^{b-1}) -1])$ with $c = \ln(3b -2)/b$ 
are also compatible with the derivative expansion \cite{opt_rg}. Their 
disadvantage is that no analytic form can be derived for RG equations 
neither in LPA nor beyond. Thus, the momentum integral in \eq{erg} has to
be performed numerically, and consequently, the dependence of the results 
on the upper bound of the numerical integration has to be considered.

The momentum integral of Eq.~\eq{erg} can be performed analytically using 
the power-law type regulator \cite{Mo1994}
\beq
\label{pow}
r_{\mr{pow}}(y) = \frac{c}{y^b }
\eeq 
at least for $b=1$ and $b=2$ in LPA. Again $c=1$ is a typical choice.
The power-law regulator is compatible with the derivative expansion (for 
any $b\geq 1$) but its disadvantage is that it is not ultraviolet (UV) safe for 
$b=1$ (at least not in all dimensions). One has to note that analyticity is lost 
beyond LPA. Therefore, similarly to the exponential type regulators, the 
dependence of the results on the upper bound of the numerical integration 
has to be considered.

Problems related to UV safety and the upper bound of the momentum 
integral can be handled by the optimized regulator function \cite{opt_rg}
\beq
\label{opt}
r_{\mr{opt}}(y) = \left(\frac{1}{y} -1\right) \theta(1-y)
\eeq 
which is a continuous function with compact support, thus the upper bound 
of the momentum integral in \eq{erg} is well-defined. A more general form
of the optimized regulator reads
\beq
\label{opt_gen}
r_{\mr{opt}}^{\mr{gen}}(y) = c \left(\frac{1}{y^b} -1\right) \theta(1-y)
\eeq 
which was discussed in detail in the context of optimization through the 
principle of minimal sensitivity \cite{minimal_sens}. Furthermore, the 
momentum integral can be performed analytically in all dimensions in LPA 
and also if the wave function renormalization is included. Moreover, it was also 
shown that in LPA, the optimized regulator and the Polchinski RG \cite{Po1984} 
equation provides us the best results (closest to the exact ones) for the critical 
exponents of the O$(N)$ symmetric scalar field theory in $d=3$ dimensions 
\cite{litim_o(n)}. This equivalence between the optimized and the Polchinski
flows in LPA is the consequence of the fact that the optimized functional RG 
can be mapped by a suitable Legendre transformation to the Polchinski one
in LPA \cite{Mo2005} but this mapping does not hold beyond LPA. It was also 
shown \cite{opt_rg} that the regulator \eq{opt} is a simple solution of the general 
criterion for optimization (see (5.10) of \cite{opt_func}) in LPA. Although, the 
regulator \eq{opt} is a continuous function but it is not differentiable and it was 
shown that it does not support the derivative expansion beyond second order. 
Indeed, it was argued in e.g. Ref. \cite{opt_func} that optimization has to meet 
the necessary condition of differentiability.

\section{The CSS regulator function}
\label{sec_cdcs}
Therefore, an appropriately chosen regulator which is a smooth function with
compact support (it has derivatives of all orders and has a finite range) can 
handle problems related to UV safety and the upper bound of the momentum 
integration in all order of the derivative expansion. In this work we give an example 
for a compactly supported smooth (CSS) regulator which has the following general 
form 
\beq
\label{css}
r_{\mr{css}}(y) = \frac{c_1}{\exp[c_2 y^{b}/(1-y^{b})] -1}  \theta(1-y)
\eeq 
with parameters $c_1$, $c_2$ and $b\geq 1$. Using the normalization 
$r_{\mr{css}}(y_0) \equiv 1$ the CSS regulator reduces to
\beq
\label{css_norm}
r_{\mr{css}}(y) = \frac{\exp[c y_0^{b}/(1-y_0^{b})] -1}{\exp[c y^{b}/(1-y^{b})] -1}  
\theta(1-y).
\eeq 
The regulator function \eq{css_norm} becomes exactly zero at $y = 1$ and
all derivatives of \eq{css_norm} are continuous everywhere. 

It is important to note here a similar problem of nuclear physics. Nuclear states are 
often described by using single-particle basis states which are eigenstates of 
single-particle Hamiltonian with phenomenological nuclear potential of strictly 
finite range (SFR) character \cite{darsl}. SFR potentials are zero at and beyond
a finite distance. The most often used spherical potential, the Wood-Saxon 
potential becomes zero only at infinity, therefore, one has to cut the tail of this 
potential if one solves the Shroedinger equation numerically. The eigenstates 
however sometimes do depend on the cut-off radius \cite{sv}. In order to get rid 
off this dependence on the cut-off radius of the Wood-Saxon form, the so called 
Salamon-Vertse (SV) potential was proposed \cite{sv}. The SV potential becomes 
zero at a finite distance smoothly, moreover the SV form can be differentiated 
any times for non-zero distance. The SV potential is a linear combination of the 
function $f(r,\rho)=-e^{\frac{r^2}{r^2-\rho^2}}\theta(1-\rho)$ and its first derivative 
with respect to the radial distance $r$. The derivative term was added to make
the SV potential be similar to the shape of the Wood-Saxon potential for heavy 
nuclei \cite{rasave,darsl}. For light nuclei one can safely use only the first term 
of the SV potential \cite{petsave}. This term in a transformed form was used as a 
weight function  
\beq
\label{weight}
w(x)\sim \theta(1-|x|)~e^{\frac{1}{x^2-1}}
\eeq
for having a finite range smoothing function for calculating the shell correction 
for weakly bound nuclei \cite{sakrve}. It is clear that 
\beq
\label{smooth}
\frac{d^{n} w(x)}{d x^n}=0 \quad \textrm {for}~ |x| \ge 1~ \textrm{and for}~ n=0,1,2,..~.
\eeq
Similar effect can be achived by using a class of functions satisfying the latter 
condition in \eq{smooth}. The present form of the CSS regulator falls into this 
class and it can be obtained from the SV potential. Similarly, the exponential 
regulator \eq{exp} is related to the Wood-Saxon potential.

In order to consider the criterion for optimization let us take the limit
\beq
\label{css_limit}
\lim_{c\to 0} r_{\mr{css}}(y) = \frac{y_0^b}{1- y_0^b}  
\left(\frac{1}{y^b} -1\right) \theta(1-y)
\eeq 
which demonstrates that the CSS regulator \eq{css_norm} recovers the 
generalized form of the optimized regulator \eq{opt_gen} and also shows that
for the particular choice $y_0 =1/2$ and $b=1$ the specified CSS regulator of 
the form
\beq
\label{css_spec}
r_{\mr{css}}^{\mr{spec}}(y) = \frac{\exp( c) -1}{\exp[c y/(1-y)] -1}  
\theta(1-y)
\eeq 
recovers the optimized one \eq{opt} in the limit $c \to 0$. Thus, for small 
enough value for the parameter $c$, the specified CSS regulator 
\eq{css_spec} produces results closer to the those obtained by the 
optimized one \eq{opt} (the smaller the parameter $c$ the closer the 
critical exponents are). Let us note, however, that if $c$ is closer to zero
higher derivates of \eq{css_spec} have sharp oscillatory peaks near $y=1$, 
thus the usage of the CSS regulator \eq{css_spec} in the limit $c\to 0$ 
requires careful numerical treatment at higher order of the derivative 
expansion. In case of an arbitrary value for $c$, the parameters $y_0$ 
and $b$ have to be redefined and the optimal choice can be done by 
using the criterion (5.10) of \cite{opt_func}. In general one finds $y_0( c)$ 
and $b( c)$ with the conditions $y_0( c\to 0)=1/2$ and $b( c\to 0)=1$. Let us 
note that the general criterion of optimization apart from (5.10) of \cite{opt_func}, 
requires a supplementary constraint related to differentiabity, see (8.42) of 
\cite{opt_func}.  It is illustrative to consider the case $y_0 = 1/2$, $b=1$ when 
these conditions can only be fulfilled by the specified CSS regulator if $c\to 0$. 
For $c \neq 0$ the determination of the optimized choice for $y_0$ and $b$ 
can only be done numerically in case of the CSS regulator which is not
investigated in this work.

Let us rewrite the CSS regulator in a more general form
\beq
\label{css_gen}
r_{\mr{css}}^{\mr{gen}}(y) = \frac{\exp[c y_0^{b}/(f-h y_0^{b})] -1}{\exp[c y^{b}/(f -h y^{b})] -1}  
\theta(f-h y^b)
\eeq 
where two new parameters $f,h$ are introduced. If one takes the following limits
\bea
\label{css_gen_limit}
\lim_{f\to \infty} r_{\mr{css}}^{\mr{gen}}(y) &=& \frac{y_0^b}{y^b}, \\ 
\lim_{h\to 0,c\to f} r_{\mr{css}}^{\mr{gen}}(y) &=& 
\frac{\exp[y_0^b]-1}{\exp[y^b]-1}
\eea
the generalized CSS regulator \eq{css_gen} reduces to the power-law \eq{pow}
and to the exponential \eq{exp} regulators. Thus the the generalized CSS
regulator \eq{css_gen} can be considered as a prototype regulator function 
which recovers all major types of regulator functions in appropriate limits.

Finally, let us note that a smooth regulator function with compact support 
has already been introduced in \cite{reuter} and it reads
\bea
\label{reuter}
r(y) =  \frac{1}{y} \exp\left[\frac{1}{y - c} \exp\left[\frac{1}{b - y}\right]\right]
\theta(c-y) \theta(y-b)  +  \frac{1}{y} \theta(b-y).
\eea 
Similarly to the CSS regulator \eq{css_gen}  it has a finite range thus it can 
handle problems related to the upper bound of the momentum integral. However,
it has an important disadvantage, namely that the regulator \eq{reuter} in its present 
form is not suitable to recover the optimized one \eq{opt}. For example, one can try 
to take the limits $c\to 1$, $b\to 0$ which result in a mixed type of regulator. 

Therefore, comparing the two compact regulators \eq{css_spec} and \eq{reuter}, 
only the CSS regulator provides us a scheme to approximate a regulator which 
fulfills the general criterion for optimization \cite{opt_func} at any order of the 
derivative expansion. The usage of the CSS regulator at higher order of the 
derivative expansion requires considerable numerical efforts for small value of 
$c$ due to the sharp oscillatory peaks of higher derivates of \eq{css_spec} near 
$y=1$ but it is differentiable for $c\neq 0$, hence, it represents an approximation 
scheme to the optimized regulator in a sense of \cite{opt_func} at all orders of the 
derivative expansion.

\section{Bosonized QED$_2$ and the CSS regulator}
\label{sec_qed2}
In order to test the specified CSS regulator function \eq{css_spec} let us study 
the critical behavior of the bosonized QED$_2$ which is the specific form of the 
massive sine-Gordon (MSG) model whose Lagrangian density is written as 
\cite{qed2}
\begin{equation}
\label{msg}
{\cal{L}}_{\mathrm{MSG}} = \hf (\partial_{\mu} \varphi)^2
+ \hf M^2 \varphi^2
+ u \cos (\beta \varphi)
\end{equation}
with $\beta^2= 4\pi$. The MSG model has two phases. The Ising-type phase 
transition \cite{dmrg_critical} is controlled by the dimensionless quantity 
$u/M^2$ which separates the confining and the half-asymptotic phases of the 
corresponding fermionic model. The critical ratio which separates the phases 
of the model has been calculated by the density matrix RG method for the fermionic 
model and the most accurate result for the critical ratio is in the range \cite{dmrg_critical}
\bea
\label{exact_ratio}
\left[\frac{u}{M^2}\right]_c  
\in [0.156,0.168].
\eea
In the framework of functional RG in LPA the closest result to \eq{exact_ratio}, i.e.
the best result for the critical ratio reads $[u/M^2]_c = 2/(4 \pi) \approx 0.15915$. It 
can be determined by analytic considerations based on the infrared (IR) limit  of the 
propagator, $\lim_{k\to 0} (k^2 + V''_k(\varphi)) = 0$ where $V_k(\varphi)$ is the 
blocked scaling potential which contains the mass term and all the higher harmonics 
generated by RG transformations \cite{scheme_sg}. This result was reproduced by 
the optimized regulator \eq{opt} and also by the power-law type one \eq{pow} with 
$b=2$ \cite{qed2}. However, if one considers the single Fourier-mode approximation
(where $V_k(\varphi)$ contains the mass term and only a single cosine) the analytic 
result based on the IR behavior of the propagator gives 
$[u/M^2]_c = 1/(4 \pi) \approx 0.07957$ \cite{scheme_sg}. In this case only the 
optimized regulator \eq{opt} was able to produce a ratio $[u/M^2]_c = 0.07964$ 
closer to the analytic one \cite{qed2}. For example, RG flows obtained by power-law 
type regulators run into a singularity and stop at some finite momentum scale and 
the determination of the critical ratio was not possible \cite{qed2}. Therefore, the 
usage of the single Fourier mode approximation provides us a tool to consider the 
convergence properties of the regulator functions.

In \fig{fig1} the phase structure of the single-frequency MSG model \eq{msg} 
is shown which is obtained by the functional RG equation derived for the 
dimensionless blocked potential ($\tilde V_k = k^{-2} V_k$) for $d=2$ 
dimensions in LPA
\bea
(2+k\partial_k) {\tilde V}_k(\varphi) = - \frac{1}{4\pi} \int_0^\infty dy 
\frac{y^2 \frac{dr}{dy}}{(1+r)y + {\tilde V''}_k(\varphi)}
\eea
using the specified CSS regulator \eq{css_spec} with $c=0.1$. The tilde 
superscript denotes the dimensionless couplings, ${\tilde M}^2 = k^{-2} M^2$ 
and ${\tilde u}_k = k^{-2} u_k$.
%
%
\FIGURE{
\epsfig{file=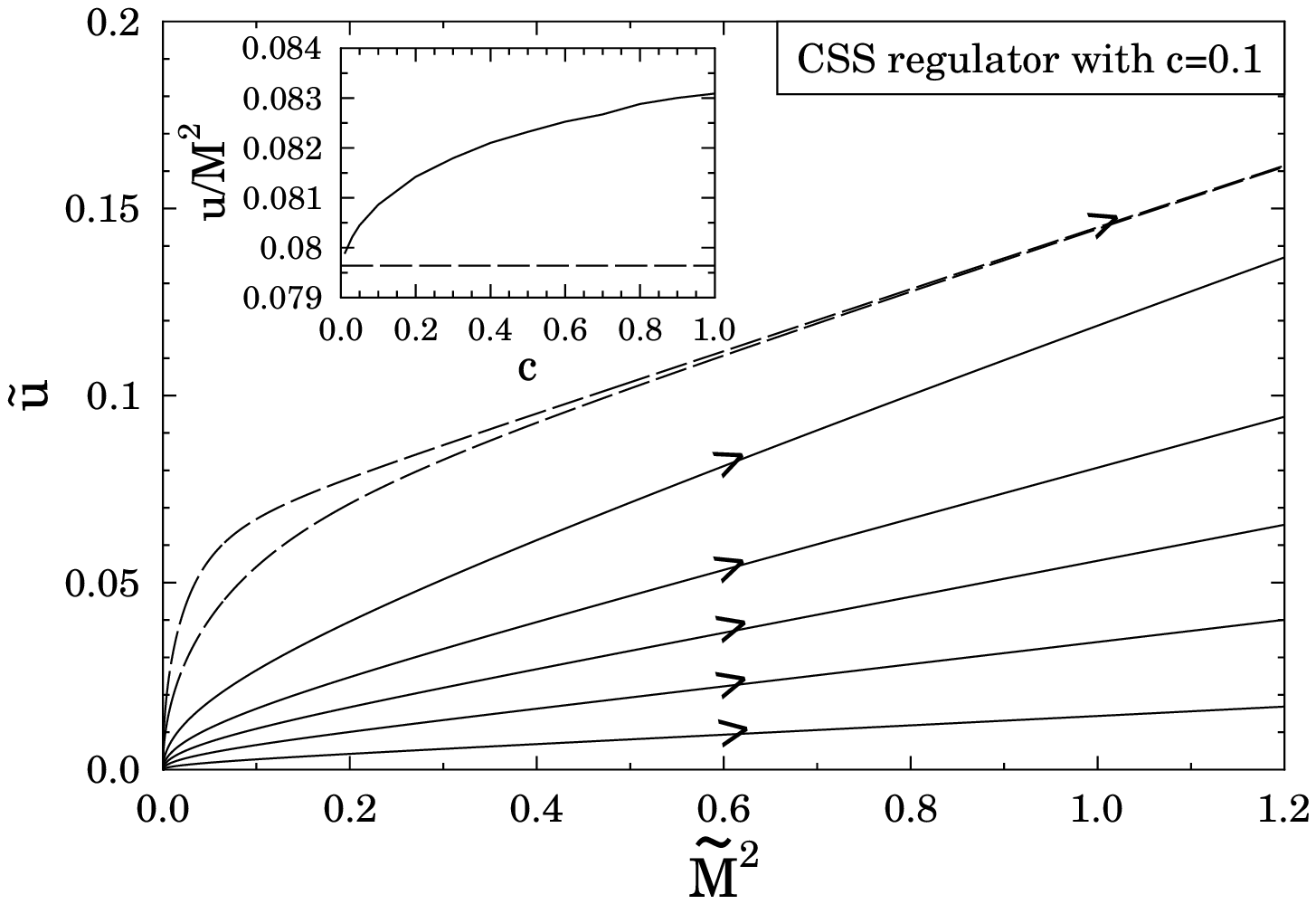,width=10 cm}
\caption{Phase diagram of the MSG model for $\beta^2 =4\pi$ is obtained by the usage 
of the CSS regulator \eq{css_spec} with $c=0.1$. RG trajectories (dashed lines) of the 
broken symmetric phase merge into a single one and the critical ratio of the model 
is determined by its slope in the IR limit. For example, $[{u}/{M^2}]_c = 0.08086$ for 
$c=0.1$ and $[{u}/{M^2}]_c = 0.07987$ for $c=0.01$. The arrows indicate the direction 
of the flow. The inset shows the dependence of the critical ratio on the parameter c 
of the specified CSS regulator \eq{css_spec} which tends to that obtained by the 
optimized regulator (horizontal dashed line).} 
\label{fig1}
}
Dashed lines correspond to RG trajectories in the broken symmetric phase 
which merge into a single trajectory in the IR limit and its slope defines the critical 
ratio. For example, $[u/M^2]_c = 0.08086$ for $c=0.1$ and $[{u}/{M^2}]_c = 0.07987$
for $c=0.01$. Let us first note that the phase structure shown in \fig{fig1} 
is almost identical to that of obtained by the optimized regulator \cite{qed2}. The inset 
of the figure shows the dependence of the critical ratio on the parameter c of the CSS 
regulator function. In the limit $c\to 0$ the critical ratio tends to that obtained by the 
optimized regulator $[u/M^2]_c = 0.07964$. Thus, it also demonstrates that the 
specified CSS regulator \eq{css_spec} reduces to the optimized one \eq{opt} in the 
limit $c\to 0$. Finally, let us note that the specified CSS regulator \eq{css_spec} has 
good convergence properties since no singularity appears in the RG flow before the 
RG trajectories merge into a single one in the broken symmetric phase similarly to 
the optimized regulator and contrary to e.g. the power-law regulator with $b=1,2$ 
\cite{qed2}.

\section{Sine-Gordon model beyond LPA and the CSS regulator}
\label{sec_sg_z}
The CSS regulator is potentially interesting for approximations beyond LPA, 
therefore, a computation including the wave function renormalization is discussed
in this section. Sine-Gordon type models are good candidates for a simple RG 
study beyond LPA because no field-dependence is required for the wave function 
renormalization (contrary to O(N) scalar theories where the field-independent 
wave function renormalization has no RG evolution, thus field-dependence is 
needed there). The MSG model \eq{msg} considered in the previous section has 
two phases in $d=2$ dimensions, so it has a non-trivial phase structure but the 
bosonization rules are violated if a cut-off dependent wave function renormalization 
is taken into account \cite{qed2}. Thus, one cannot compare directly the results of 
the RG study of the MSG model to those of the QED$_2$. The pure sine-Gordon
(SG) model without a mass term defined by the Euclidean action,
\beq
\label{sg}
S = 
\int d^d x \left[\hf (\partial_\mu\varphi)^2 + u\cos(\beta \varphi)\right],
\eeq
has a trivial single phase for $d>2$ \cite{cg_sg} but it undergoes a topological
phase transition for $d=2$ \cite{scheme_sg,sg_KTB}. However, the critical value
($\beta^2_c = 8\pi$) which separates the phases of the SG model in $d=2$ dimensions 
was found to be scheme-independent \cite{scheme_sg}. Therefore, in order to test 
the CSS regulator in the framework of SG type models the best choice is an RG study 
of the SG model for $d<2$ dimensions where the position of the non-trivial saddle 
point which separates the two phases is scheme-dependent \cite{cg_sg}. Indeed, 
the phase structure of the SG model for $d<2$ dimensions has been investigated in 
\cite{cg_sg} by solving the RG flow equations derived for the dimensionful couplings 
($u_k$ and $z_k = 1/\beta^2$),
\bea
\label{exact_u}
k\partial_k u_k =
\int _p \frac{k\partial_k R_k}{k^{2-d} u_k}
\left(\frac{P-\sqrt{P^2-(k^{2-d} u_k)^2}}{\sqrt{P^2-(k^{2-d} u_k)^2}}\right),\\
\label{exact_z}
k\partial_k z_k = \int_p \frac{k\partial_k R_k}{2}
\biggl[
\frac{-(k^{2-d}u_k)^2P(\partial_{p^2}P+\frac{2}{d}p^2\partial_{p^2}^2P)}
{[P^2-(k^{2-d}u_k)^2]^{5/2}}\nn
+\frac{(k^{2-d}u_k)^2 p^2 (\partial_{p^2}P)^2(4P^2+(k^{2-d}u_k)^2)}
{d \, [P^2-(k^{2-d} u_k)^2]^{7/2}}
\biggr] ,
\eea
where $P = z_k k^{2-d} p^2+R_k$ and $\int_p = \int dp \, p^{d-1} \Omega_d/(2\pi)^d$  
with the d-dimensional solid angle $\Omega_d$. The phase diagram was obtained 
by the power-law \eq{pow} RG with $b=2$ and was plotted in Fig.1 of  \cite{cg_sg} 
(for $d=1$). The position of the saddle point is scheme-dependent and for $b=2$ it 
is given by $\bar{u}_\star = 0.57$, $1/{\tilde z_\star} = 7.95$ where the normalized 
coupling $\bar{u}_k \equiv  k^{2-d} u_k/\bar{k} = k^2{\tilde u}/\bar{k}$ is defined by 
the dimensionless one $\tilde u_k$ and $\bar{k} = \min_{p^2} P$.

Let us map out the phase structure of the SG model for $d<2$ dimension by means
of the CSS RG.  Inserting \eq{css_spec} into \eq{exact_u} and \eq{exact_z} one obtains 
the phase diagram (for $d=1$) plotted in \fig{fig2} which is similar to that of obtained by 
the power-law RG with $b=2$.
%
%
\FIGURE{
\epsfig{file=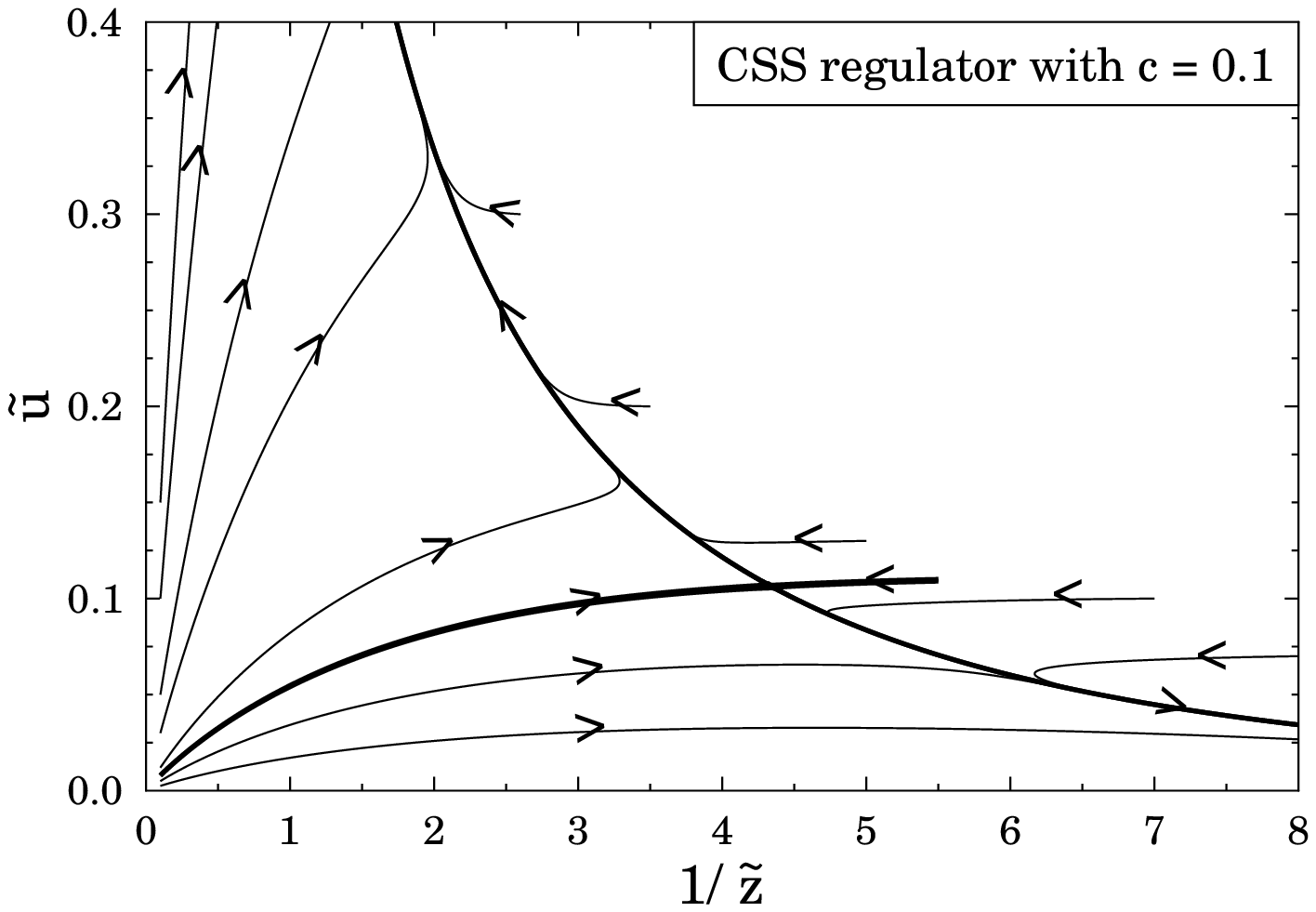,width=10 cm}
\caption{Phase diagram of the SG model for $d=1$ dimension is obtained by the 
usage of the CSS regulator \eq{css_spec} with $c=0.1$ including the wave function 
renormalization. Similar RG flow can be drawn for $1<d<2$. Arrows indicate the 
direction of the flow.} 
\label{fig2}
}
The two attractive IR fixed points  ($\tilde u_\star = 1.05$, $1/{\tilde z_\star} = 0$ and 
$\tilde u_\star = 0$, $1/{\tilde z_\star} = \infty$) indicate two phases. The coordinates 
of the non-trivial saddle point which separates the phases read as $\tilde u_\star = 0.106$, 
$1/{\tilde z_\star} = 4.34$. Using the normalized coupling $\bar{u}_k = k^2{\tilde u}/\bar{k}$ 
one obtains $\bar{u}_\star = 0.47$ which is close to that of given by the power-law RG with 
$b=2$. Thus, the CSS RG provides us reliable results beyond LPA, too.

Although the detailed analysis of the regulator dependence of the above result is 
beyond the scope of the present work, let us briefly discuss the scheme-dependence 
of the RG study of the SG model for dimensions $d<1$ beyond LPA. Both for fractal 
dimensions $1<d<2$ and for $d=1$, the non-trivial saddle point appears in the RG flow 
of the SG model. However, there is an important difference between the two cases. In 
one-dimension as a consequence of the equivalence between quantum field theory and 
quantum mechanics a symmetry cannot be broken spontaneously due to the tunneling 
effect. Thus, for one-dimensional quantum field theoric models the (spontaneously) 
broken phase should vanish if their phase structure have been determined without 
using approximations. Therefore, the requirement of the absence of the broken phase 
in case of the non-approximated RG flow can be used to optimize the RG 
scheme-dependence of the approximated one \cite{opt_d1_d2}. The broken phase 
vanishes if the saddle point coincides with the non-trivial IR fixed point found at 
$\tilde u_\star = 1.05$, $1/{\tilde z_\star} = 0$. Thus, the distance between the saddle 
point and the non-trivial IR fixed point can be used to optimize the RG equations, i.e. 
the better the RG scheme is the closer the fixed points are \cite{opt_d1_d2}. We note that 
one has to use appropriately normalized couplings such as $\bar{u}_k = k^2{\tilde u}/\bar{k}$
and $\bar z_k \equiv (8\pi) \tilde z_k $. Then, the distance between the saddle point 
and the non-trivial IR fixed point should be minimized in order to obtain the optimal
choice for the parameters of a given regulator function. In Ref. \cite{opt_d1_d2} this 
new type of optimization scenario was tested first for the power-law regulator and
the known results were recovered. Then the optimization of the RG flow obtained by 
generalized CSS regulator function \eq{css_gen} was performed. It has importance 
since the generalized CSS regulator is a prototype regulator which recovers 
all major type of regulator functions in appropriate limits, thus, its optimization can 
produces us the best choice among the class of regulator functions. For example,
it can be shown that by the fine tuning of the parameters of the CSS regulator it is 
possible to produce better results (smaller distance) then by the power-law type 
regulator \cite{opt_d1_d2}.

\section{Summary}
\label{sec_sum}
In this work an example was given for a compactly supported smooth (CSS) 
regulator function. Similarly to the optimized (in a sense of \cite{opt_rg,opt_func}) 
regulator it has a finite range, hence, the upper bound of the momentum integral 
of the functional RG equation is well-defined in numerical treatments and it is UV 
safe. Since the CSS regulator is a function of class $C^\infty$ its advantage is that 
it has derivatives of all orders in contrary to the optimized regulator which is 
continuous but not differentiable. This has important consequences on the 
applicability of the CSS regulator beyond the second order of the derivative 
expansion. It was also shown that in the limit $c\to 0$ the specified CSS regulator 
reduces to the optimized one \eq{opt}, therefore, the smaller the parameter 
$c$ the closer the results obtained by the two regulators are. Moreover, it was 
also shown that the generalized form of the CSS regulator can be considered 
as a prototype regulator which reduces to all major type of regulator functions 
(exponential, power-law) in appropriate limits. Although, the usage of the CSS 
regulator at higher order of the derivative expansion requires considerable 
numerical efforts for small value of $c$ due to the sharp oscillatory peaks of 
higher derivates near $y=1$ but it is differentiable for $c\neq 0$, hence, it 
represents an approximation scheme to the optimized regulator in a sense of 
\cite{opt_func} at all orders of the derivative expansion. This was demonstrated 
by considering the critical behavior of the bosonized QED$_2$ in the local 
potential approximation. The CSS regulator has been tested beyond the local 
potential approximation in the framework  of the sine-Gordon scalar theory for 
$d<2$ dimensions. A similar smoothing problem of nuclear physics was also 
discussed.

\section*{Acknowledgement}
This research was supported by the T\'AMOP 4.2.1./B-09/1/KONV-2010-0007 project.
Fruitful discussions with Tam\'as Vertse, P\'eter Salamon and Andr\'as Kruppa on the
smoothing problem and the usage of the Salamon-Vertse potential in nuclear physics 
are warmly acknowledged. The author thanks discussions on the possible generalization
of smooth functions with compact support for P\'eter Salamon. Furthermore, the author  
acknowledges useful discussions with Bertrand Delamotte on the numerical treatment 
of the CSS regulator in particular the oscillatory behavior of its higher derivatives. 
Daniel Litim is warmly acknowledged for comments on optimization and also for paying 
the attention of the author on the papers by O. Luscher and M. Reuter where a compact 
regulator has also been discussed.

\end{document}